# Optimize Energy Consumption of Wireless Sensor Networks by using modified Ant Colony Optimization ACO

Yasameen sajid Razooqi[1,2,*], Muntasir Al-Asfoor[2,3], Mohammed Hamzah Abed[2,4]

[1] Al-Qadisiyah Education Directorate Al Diwaneyah, Iraq
[2] Computer Science Department, Al Qadidiyah University, 58002 Al Diwaneyah, Iraq
[3] Buckinghamshire New University, Queen Alexandra Rd, High Wycombe HP11, UK
[4] Department of Telecommunications and Media Informatics, Budapest University of Technology and Economics, Magyar tudósok körútja 2, 1117 Budapest, Hungary
*e-mail: yasameen.sajid@qu.edu.iq

**Abstract:** Routing represents a pivotal concern in the context of Wireless Sensor Networks (WSN) owing to its divergence from traditional network routing paradigms. The inherent dynamism of the WSN environment, coupled with the scarcity of available resources, engenders considerable challenges for industry and academia alike in devising efficient routing strategies. Addressing these challenges, a viable recourse lies in applying heuristic search methodologies to ascertain the most optimal path in WSNs. Ant Colony Optimization (ACO) is a well-established heuristic algorithm that has demonstrated notable advancements in routing contexts. This paper introduces a modify routing protocols based on Ant colony optimization. In these protocols, we incorporate the inverse of the distance between nodes and their neighbours in the probability equations of ACO along with considering pheromone levels and residual energy. These formulation modifications facilitate the selection of the most suitable candidate for the subsequent hop, effectively minimizing the average energy consumption across all nodes in each iteration. Furthermore, in this protocol, we iteratively fine-tune ACO's parameter values based on the outcomes of several experimental trials. The experimental analysis is conducted through a diverse set of network topologies, and the results are subjected to comparison against well-established ACO algorithm and routing protocols. The efficacy of the proposed protocol is assessed based on various performance metrics, encompassing throughput, energy consumption, network lifetime, energy consumption, the extent of data transferred over the network, and the length of paths traversed by packets. These metrics collectively provide a comprehensive evaluation of the performance attainments of the routing protocols.

*Keywords:*   *WSN ; Energy ; Optimization ; ACO*

## I. INTRODUCTION

Wireless Sensor Networks (WSNs) encompass a set of miniature devices called sensors, designed to gather data from the observed environment and transmit it to a central base station known as the Sink [1]. Based on the specific application or purpose, these sensors can collect diverse data types, including temperature, noise, and pressure. The Sink processes the received data and may forward it to remote users or cloud platforms, facilitating integration with the Internet of Things (IoT). The IoT finds extensive applications in various domains, such as forest and ocean monitoring, nuclear reactor area management, and healthcare [2]. Figure 1 depicts the integration of WSN with IoT. Each sensor comprises three principal components: the sensing unit responsible for data acquisition, the processing unit housing storage memory, and the communication unit for data transmission and reception. Additional components, such as GPS and mobilizers for mobility management in the case of mobile sensors, may also be incorporated, albeit at the expense of higher energy consumption [3]. To address energy constraints, solar cells can be integrated as an auxiliary power source [4]. However, the primary power source for sensors remains batteries, posing a significant challenge in WSNs, as these batteries are typically non-rechargeable and non-replaceable due to the sensors' remote and hazardous deployment locations, coupled with their small size and low cost [5]. The communication phase incurs the highest energy consumption, surpassing that of sensing and processing. Given the extensive deployment of sensors over vast geographical areas, manual management becomes impractical, necessitating dynamic routing techniques in WSNs to minimize energy consumption during communication. Achieving this objective involves leveraging artificial intelligence techniques and distributed algorithms to learn the network topology and determine energy-efficient paths [6][7]. This study proposes an enhancement of flat routing protocols for WSNs by modification of ACO algorithm.

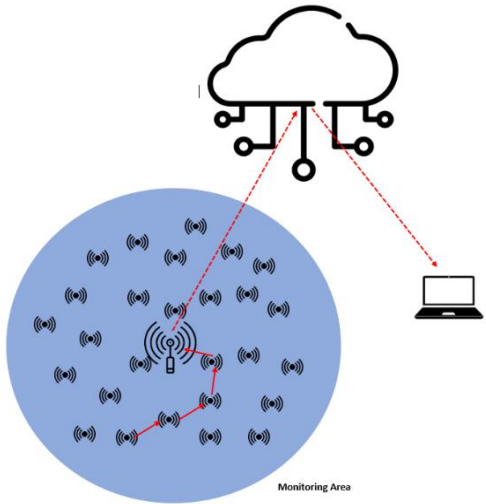

*Figure 1* Wireless Sensor Network components and their integration with IoT

Routing in Wireless Sensor Networks (WSNs) presents several formidable challenges, including those about energy consumption, communication range, and the establishment of peer-to-peer (P2P) communications. In P2P networks, all devices are interconnected with equal permissions and shared responsibilities for data processing. Unlike client and server networks, no specific devices in P2P networks are designated solely for sending or receiving data [8]. Each device possesses the same rights as its peers and can fulfil analogous functions. Consequently, in WSNs, each sensor must judiciously select the most suitable neighbour within its communication range, facilitating the formation of efficient multi-hop paths leading to the sink node. The goal is to minimize the distance traversed and, thus, conserve energy during data transmission. Figure 2. visually depicts the sensing and communication ranges of the sensors.

Routing in Wireless Sensor Networks (WSNs) determines network lifetime and overall performance. Given the reliance of sensors on battery power and the significant energy consumption in communication, efficient routing management is essential to reduce energy consumption while maintaining high throughput. This study proposes leveraging Artificial Intelligence (AI) algorithms to determine the next hop or candidate neighbor for packet transmission, a key distinction of WSN routing from other network types. Each sensor possesses limited knowledge of its immediate neighbors within its communication range rather than the entire network, necessitating novel approaches for flat routing protocols that do not rely on central nodes. To address this challenge, we introduce two intelligent routing protocols. In this protocol, we utilities an enhanced (ACO) algorithm to select the next-node while searching for the sink (destination). The modification involves to perform thorough testing and analysis of various network topologies to determine the optimal parameter values ($\alpha$, $\beta$, $\rho$, $\gamma$, and Q) for ACO in the flat routing protocol within WSNs. To mitigate energy consumption, we propose reducing the communication range of the sensors to prevent packets from being transmitted across long distances, a strategy based on the energy model discussed later. Additionally, we aim to minimize the Time-to-Live (TTL) value (number of hops) based on the network size. Furthermore, to address the issue of fast node depletion in flat protocols, we propose using super sensors near the sink to handle high loads efficiently, mitigating the risk of nodes dying prematurely. These proposed strategies are expected to improve the efficiency and performance of flat routing protocols in WSNs, leading to prolonged network lifetime and enhanced overall operation.

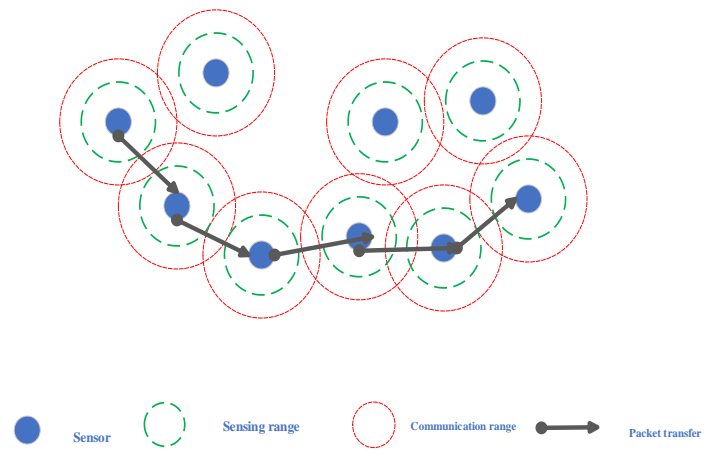

*Figure 2* the packet routing in WSN based on communication range

This paper is organized as follows: Section two present the related work, Section three the materials and proposed models of this work, followed by a result analysis illustrated in Section four. Section five presents a conclusion of all the concepts and the future direction in this research field.

## II. RELATED WORK

This section presents a review of various research studies focused on reducing energy consumption in Wireless Sensor Networks (WSN) through hierarchical and flat routing approaches. In one study [8], the researchers employed Capsule Neural Network to develop a learning model for identifying suitable cluster heads within the cluster based on identity records. They also utilized shortest path selection to identify forward nodes outside the clusters, reducing energy consumption. In another research effort [9], gravitational force and fuzzy rules were integrated to construct clusters and manage routing, improving network lifetime. Where the fuzzy logic has been used to select appropriate nodes as cluster heads. In another hand, the Energy-efficient Scalable Clustering Protocol [10] addressed distances between and within clusters to generate equiponderant clusters. The Particle Swarm Optimization technology based on Dragonfly's algorithm (DA–PSO) was employed to select a cluster heads, along with a new energy-efficient fitness function for optimal CH selection. However, hierarchical routing protocols suffer from inefficiencies in cluster head selection and frequent cluster head changes during the network's lifetime, along with the rapid depletion of cluster heads due to their dual roles of collecting sensing data and forwarding it to the sink.

Another study [11] employed Ant Colony Optimization (ACO) to select the next hop based on residual energy and pheromone as heuristic functions.

It is important to note that the heuristic function based solely on save the residual energy in [12][13] did not guarantee to reach the sink in fewer hops, resulting in decreased throughput despite high energy consumption. One significant limitation of flat routing protocols is the rapid energy depletion of sensors near the sink due to the heavy load of transmitting sensing data from other sensors. In contrast, the proposed protocol addresses this limitation by utilizing solar cells as a power supply for the sensors near the sink and leveraging methods based on artificial intelligence to find the minimum length of the success path, thereby enhancing energy efficiency.

## III. MATERIALS AND METHODS

### 1. Network setting

Two distinct models govern the distribution of sensors within the monitoring area, the pre-planned mode, wherein the monitoring area is readily accessible for manual sensor deployment and management, resulting in efficient coverage with a reduced number of sensors; and the ad-hoc mode, which assumes greater significance in challenging or distant environments where manual access is impractical. In the ad-hoc mode, a larger number of sensors are deployed randomly to achieve the required coverage [14][15][16]. The latter distribution model was adopted during the network's construction phase, involving the random deployment of N sensors within a square area. The network configurations encompassed three setups, each varying in the number of nodes and network area, specifically: 80 nodes in a 100*100meter region, 160 nodes in a 200*200-meter region, and 240 nodes in a 300*300-meter region.

During the network operation, each sensor initiates communication by sending a hello packet to its neighbouring sensors, conveying essential information such as ID, location, and energy status. Data transfer within the network was categorized into three modes: event-driven, where data transmission commences upon the occurrence of specific events; time-driven, wherein sensed data is transmitted to the sink at predetermined intervals; and query-driven, where data transmission is triggered in response to sensor requests [6]. In the proposed model, the first type, i.e., event-driven data transfer, was adopted to enable a specific group of sensors to transmit their sensing data during each operational round. To simulate this behavior, the set of sensors responsible for data transmission during each round was randomly selected.

### 2. Energy model

In this study, we assume that all sensors initially possess equal energy levels. However, over time, each sensor's energy varies due to energy depletion during the sensor operations (sensing, data processing, and communication). The energy expended on sensing and processing is minimal compared to the energy used for communication [17].

Every sensor keeps an information table that holds information about neighbouring nodes: their IDs, remaining energy levels, and positions. The Euclidean distance formula is used to calculate the distance between nodes, and any changes in energy levels are promptly updated in the table. To evaluate energy consumption during communication, we utilize the equations specified in reference [18]. Specifically, depending on the distance between sender and receiver sensors, one of two equations is applied. When the distance is below a predefined threshold "d0," the free space model is employed. On the other hand, if the distance surpasses this threshold, the multipath fading model is used to calculate the energy needed for transmitting a 1-bit packet.

$$E_T(l, d) = \begin{cases} lE_{elec} + \varepsilon_{fs}d^2 & if \ d < d_o \\ lE_{elec} + \varepsilon_{mp}d^4 & if \ d \geq d_o \end{cases} \quad (1)$$

$E_{elec}$ is the energy of the electronic circuit, while $\varepsilon_{fs}$ and $\varepsilon_{mp}$ are the energy consumed. In the receiver sensor, energy lost in receiving *l* bits is:

$$ER(l) = lE_{elec} \quad (2)$$

### 3. Ant colony optimization

Ant Colony Optimization (ACO) has been utilities in Wireless Sensor Networks (WSNs) to facilitate sensor nodes in selecting the most suitable neighbors by augmenting their knowledge with additional pheromone information integrated into the sensor table. In this study, we propose enhancement of Ant Colony Optimization for flat routing protocol, an intelligent protocol based on ACO. Initially, all connections between sensors are assigned uniform initial pheromone values. Consequently, during the first hop, sensor nodes make their next-hop selections without being influenced by pheromone concentration. However, the pheromone values are updated with each successive round, impacting the subsequent next-hop selections.

To enhance the heuristic function in the probability equation, we suggest two methods in the proposed Modified ACO. Firstly, we employ the inverse of the distance between nodes i and j to establish the heuristic function $\eta_{ij}$, favouring a higher probability for shorter distances, thereby encouraging the selection of closer sensors. Secondly, we incorporate two additional heuristic functions, along with the pheromone value, by incorporating the distance between nodes i and j and the residual energy of the neighbor sensor. Consequently, the probability equation takes the form of equation (3):

$$p_{ij}^m(t) = \left\{ \frac{[\tau_{ij}(t)]^\alpha \cdot [n_{ij}(t)]^\beta \cdot [\delta_j(t)]^\gamma}{\sum v_m \in v_{allowed}^m [\tau_{im}(t)]^\alpha \cdot [n_{im}(t)]^\beta \cdot [\delta_j(t)]^\gamma} \right.$$

$$if\ v_j\ v_{allowed}^m \quad (3)$$

Where $\eta_{ij}$ represents the inverse of the distance between nodes i and j, and $\delta_j$ denotes the difference between the initial and residual energy of node j.

To perform pheromone updates, we utilized equation 4 and 5.

$$\tau_{ij}(t + \Delta t) = (1 - \rho) \cdot \tau_{ij}(t) \quad \text{for all edge in the network} \quad (4)$$

And

$$\tau_{ij}(t + \Delta t) = \tau_{ij}(t) + \Delta\tau_{ij}(t) \quad \text{for all success paths} \quad (5)$$

The ACO's parameters (α, β, ρ, γ, and Q) were systematically adjusted after conducting rigorous testing and analysis across the three network topologies to determine the values that optimize high throughput while minimizing energy consumption, rendering ACO suitable for this particular context. Algorithm 2 shows the steps of modified ACO.

## IV. EXPERIMENT RESULT AND ANALYSIS

This section provides a comprehensive overview of the networks' attributes, the simulation environment employed, the designated parameters, and the performance evaluation results for the two proposed protocols. Additionally, we present a comparative analysis of the outcomes of these protocols concerning the performance metrics adopted for the evaluation. Furthermore, we juxtapose the results of the proposed protocols against other algorithms, which is utilities in flat routing approaches used in Wireless Sensor Networks (WSNs).

### 1. Environment simulations

In the simulation of network configuration, implementation of routing protocols, and subsequent result analysis, we utilized the Python programming language, complemented by the NetworkX package. To substantiate the efficacy of our proposed protocols, we conducted simulations on three distinct networks. These networks comprised 80 nodes in a 100*100-meter area, 160 nodes in a 200*200-meter area, and 240 nodes in a 300*300-meter area. Table 1 shows the network scenarios in our experiments. parameters of the network's scenarios.

**Table 1** Network scenarios

| parameters | Scenario 1 | Scenario 2 | Scenario 3 |
|---|---|---|---|
| Network Area | 100×100 m² | 200×200 m² | 300×300 m² |
| Base station location | (50,50) | (100,100) | (150,150) |
| Number of sensors | 80 | 160 | 240 |
| Deployment | Random -Scale free | Random - Scale free | Random - Scale free |
| Communication range | 20 | 28 | 35 |

Furthermore, Table 2 accounts for the parameters employed in the network model, energy model, and the Ant Colony Optimization (ACO) algorithm. These parameters remained consistent throughout our experimental analysis for all the protocols.

**Table 2** Experimental parameters

| Parameters | Value |
|---|---|
| Initial energy of sensors | 5 J |
| Number of sinks | 1 |
| Location of sink | Centre of area |
| Packet size | 1024 |
| $E_{elec}$ | 50 nJ/bit |
| $\mathcal{E}_{mp}$ | 0.0013 pJ/bit/m4 |
| $\mathcal{E}_{fs}$ | 10 pJ/bit/m2 |
| $d_0$ | 50 |
| α | 2 |
| β | 3 |
| Initial pheromone | 1 |
| ρ | 0.5 – 1 |
| Q | 1 |

### 2. Result and analysis

In this study, we have comprehensively assessed the simulation outcomes encompassing throughput, energy consumption, network lifetime, and success message ratio travelled by data packets. We have juxtaposed the performance of our proposed ACO-based routing algorithm, outlined in [16], with that of the conventional ACO routing protocol under identical simulation conditions, thus substantiating the efficacy of the algorithms in question. In the subsequent section, we will expose each of the aforementioned metrics, delineated to two distinct scales: the count of network operation iterations or rounds and the progressive Time-to-Live (TTL) range. Each TTL value corresponds to the network's performance following 1000 iterations of network operation. Furthermore, these metrics will be illustrated across three distinct network topologies, explicated in Table 1.

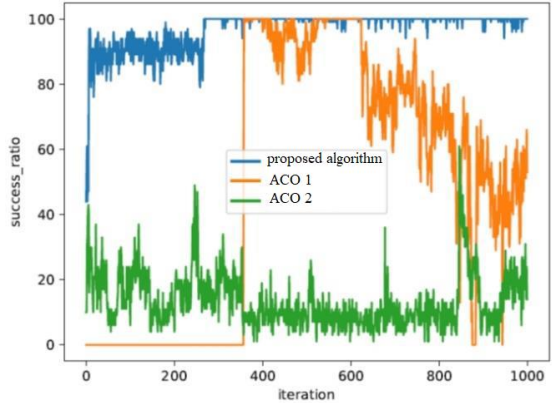

*Figure 3* Success message ratio based on 160 network nodes.

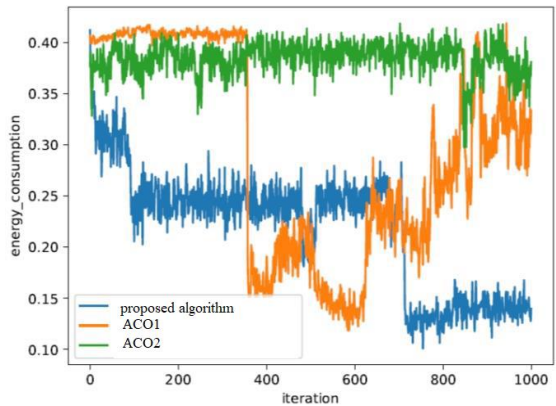

*Figure 4* Energy consumption ratio based on 160 network nodes.

The throughput metric indicates the successful delivery of packets to their intended destinations, with the success ratio employed as an illustrative measure in our calculations. The figures presented above conspicuously demonstrate the notable superiority of the ACO-based routing protocol (proposed algorithm) in achieving elevated throughput compared to both the traditional ACO protocol and the energy and randomness-based routing approaches. Figures 3 and 4 visually depict the success rates across the three network topologies during varying operational iterations and energy consumption. Additionally, Figure 5 showcases the aforementioned metrics in scenarios involving different TTL values, thereby further substantiating the superior throughput performance of our proposed ACO-based routing protocol. And figure 6 shows final comparison of Network life time. And its show the proposed ACO achieved high performance compare to the traditional methods.

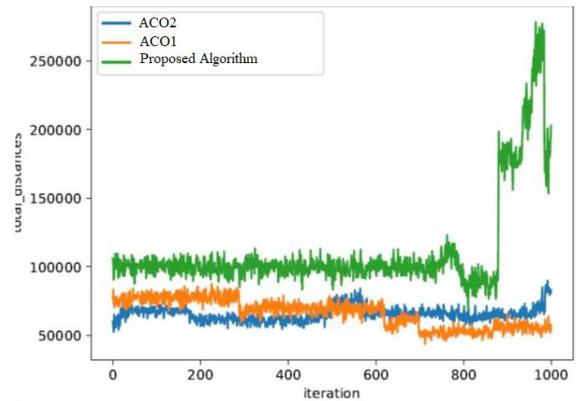

*Figure 5* TTL values over 160 network nodes

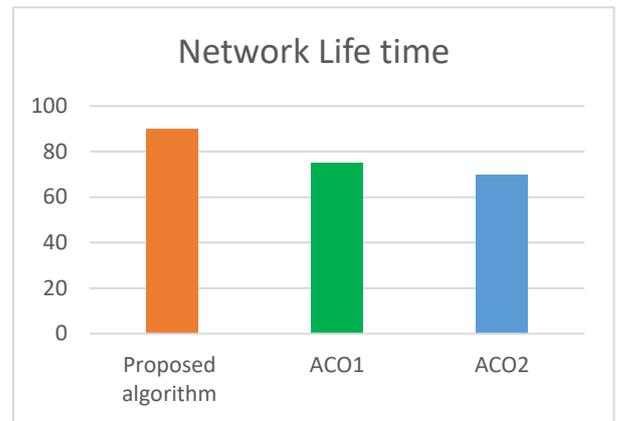

*Figure 6* Network life time

## V. CONCLUSION

In this research endeavour, we present a series of intelligent routing protocols designed to enhance the operational longevity of Wireless Sensor Networks (WSNs). Our primary objective revolves around the optimization of energy utilization within these networks. To achieve this, we leverage the heuristic search optimization approach known as Ant Colony Optimization (ACO). Our efforts encompass refining the underlying pheromone concentration equations within the ACO algorithm. Moreover, we meticulously fine-tune the pivotal ACO parameters (α, β, ρ, γ, and Q) based on an iterative process involving comprehensive testing across diverse network topologies. This meticulous parameter calibration is geared towards rendering the ACO algorithm conducive to seamless integration within the framework of flat routing protocols within WSNs. The proposed algorithm's efficacy is examined across three distinct yet controlled network scenarios. These scenarios encapsulate 80 nodes deployed within a 100x100 meter area, 160 nodes spanning a 200x200 meter expanse, and 240 nodes dispersed across a 300x300 meter terrain. Notably, all evaluated scenarios are governed by identical network and energy parameters, ensuring an equitable basis for comparison. The evaluation outcomes compellingly underscore the superior performance of the proposed modified ACO algorithm. This superiority contrasts not only with the conventional ACO algorithm as documented in [16] but also against a spectrum of alternative protocols. This ascendancy is consistently evident across all dimensions of performance metrics embraced within this comprehensive study.

## ACKNOWLEDGEMENT

We would like to thanks Al-Qadisiyah University, Buckinghamshire New University, Budapest University of technology and economics for their support.

## AUTHOR CONTRIBUTIONS

**A.Author**: Conceptualization, Experiments, Theoretical analysis.

**B. Author**: Supervision, Review and editing.

**C. Author**: Supervision, Review and editing.

## DISCLOSURE STATEMENT

The authors declare that they have no known competing financial interests or personal relationships that could have appeared to influence the work reported in this paper.

## ORCID

If the authors have ORCID identification, it must be given in this section.

**A. Author** https://orcid.org/0000-0003-2315-067X

**B. Author** https://orcid.org/0000-0002-3742-1963

**C. Author** https://orcid.org/0000-0003-4780-4252